\documentclass[aps,prd,twocolumn,superscriptaddress,nofootinbib,preprintnumbers,10pt]{revtex4-1}
\usepackage{style}

\begin{document}

\preprint{FERMILAB-PUB-26-0702-T}

\title{A Simple Dark Matter Model to Explain the LZ Event and Galactic Center Excess}

\author{Caleb Gemmell \,\orcidlink{0000-0002-6505-8559}}
\email{cgemmell2@wisc.edu}

\affiliation{Department of Physics, University of Wisconsin, Madison, WI, USA}

\affiliation{Wisconsin IceCube Particle Astrophysics Center,
University of Wisconsin, Madison, WI, USA}

\author{Dan Hooper\,\orcidlink{0000-0001-8837-4127}}
\email{dwhooper@wisc.edu}

\affiliation{Department of Physics, University of Wisconsin, Madison, WI, USA}

\affiliation{Wisconsin IceCube Particle Astrophysics Center,
University of Wisconsin, Madison, WI, USA}

\author{Gordan Krnjaic\,\orcidlink{0000-0001-7420-9577}}
\email{krnjaicg@fnal.gov}
\affiliation{Theory Division, Fermi National Accelerator Laboratory}
\affiliation{University of Chicago, Department of Astronomy and Astrophysics}
\affiliation{University of Chicago, Kavli Institute for Cosmological Physics}

\begin{abstract}
We propose that the dark matter (DM) is a pseudo-Dirac fermion with a mass of approximately $20-100$ GeV that couples inelastically to the Standard Model through a hadrophilic spin-1 mediator, $V$, the gauge boson of spontaneously broken baryon number. The two DM eigenstates, $\chi_{1}$ and $\chi_2$, acquire approximately equal abundances after thermal freeze-out and are depleted through the processes $\chi_1 \chi_1 \to VV$ and $\chi_2 \chi_2 \to VV$. Although the heavier state is unstable, for a mass splitting below the electron–positron threshold and sufficiently small mass mixing with the $Z$ boson, its lifetime can exceed the age of the universe, allowing both states to remain approximately equally abundant in the Galactic halo. For a thermal-relic annihilation cross section, this model can produce a signal that is consistent with the observed features of the longstanding Galactic Center Gamma-Ray Excess (GCE). If $\delta \sim 1000$ keV,  the same parameter space that explains the GCE can also explain the LZ event through exothermic $\chi_2 \to \chi_1$ downscattering off xenon nuclei.
\end{abstract}

\maketitle

\textbf{\textit{ Introduction:}} The  LUX-ZEPLIN  (LZ) Collaboration recently reported an intriguing event in a search for dark matter (DM) interactions in which the nuclear-recoil energy window was extended to approximately $270\,{\rm keV}$~\cite{LZ:2026axp}. In an exposure of 2.84 tonne-years, LZ observed a single event consistent with a nuclear recoil of $248\pm23\,({\rm stat})\pm23\,({\rm sys})\,{\rm keV}$ in a region where the expected background is very low. The event is difficult to explain with conventional elastic, spin-independent DM scattering, which generally predicts a spectrum concentrated at lower recoil energies, but could arise more naturally from momentum-dependent or inelastic interactions~\cite{Fan:2026kxx,Freese:2026sga,Wu:2026nhi,Yin:2026jnn,Su:2026rwz,Yamashita:2026ump,DiMauro:2026ldr,Unwin:2026rdp,McCabe:2026crm,Arcadi:2026kev,An:2026pkc,Borah:2026ris,Cabo-Almeida:2026uqw,Nguyen:2026lui,Lee:2026jxl,Langhoff:2026ujr,Borah:2026zwf,Nomura:2026qyq}. Across the models considered by LZ, the largest local significance was $3.4\sigma$, which decreases to a global significance of $2.6\sigma$ after accounting for the look-elsewhere effect. 

This event is particularly interesting in the context of the longstanding signal known as the Galactic Center Gamma-Ray Excess (GCE), observed by the Fermi Large Area Telescope (LAT)~\cite{Goodenough:2009gk,Hooper:2010mq,Hooper:2011ti,Abazajian:2012pn,Hooper:2013rwa,Gordon:2013vta,Daylan:2014rsa,Calore:2014xka,Fermi-LAT:2015sau,Cholis:2021rpp,DiMauro:2021raz,DiMauro:2026fnp}. The spectrum, morphology, and intensity of the GCE are each consistent with expectations from annihilating DM. In the case of annihilations to $b\bar{b}$, for example, the excess is well fit by DM with a mass of $m_{\chi} \sim 30-75 \, {\rm GeV}$ and that is distributed in the Inner Galaxy with an approximate profile of $\rho \propto r^{-1.2}$. The normalization of this signal requires an annihilation cross section on the order of $\langle \sigma v \rangle \sim 10^{-26} \, {\rm cm}^3/{\rm s}$, consistent with that expected for DM in the form of a generic thermal relic~\cite{PhysRevD.86.023506}. Although this signal could potentially be generated by a large population of millisecond pulsars, such a scenario would require a very large number ($\sim 10^5$) of such pulsars and for those sources to be systematically fainter
than those observed in other environments~\cite{Holst:2024fvb,Amerio:2024qor,List:2025qbx}.

Many of the DM models that could have generated the GCE have been ruled out by the null results of direct detection experiments (for reviews, see Refs.~\cite{Escudero:2016kpw,Kong:2025ccv}). There are, however, several exceptions to this conclusion, including models in which the DM annihilates to hidden-sector particles that subsequently decay into Standard Model (SM) particles~\cite{Berlin:2014pya,Hooper:2019xss,Escudero:2017yia,Hu:2025thq}, or models in which the DM interactions are mediated by a particle with pseudoscalar couplings~\cite{Berlin:2014tja,Ipek:2014gua,Berlin:2015wwa,Karwin:2016tsw,Hu:2025thq}.

In this {\it Letter}, we point out that a DM candidate in the form of a pseudo-Dirac fermion coupled to a hadrophilic gauge boson can accommodate the observed features of the GCE and could also be responsible for the event recently reported by LZ.

\medskip
\textbf{\textit{Model Description:}} In the model under consideration, we extend the SM with a new $U(1)_B$ gauge group and a corresponding gauge boson, $V$, where $B$ is baryon number. We also introduce a new Dirac fermion $\chi = (\xi, \eta^\dagger)$, where $\xi$ and $\eta$ are Weyl spinors with equal and opposite baryon charges. Following the conventions in Ref. \cite{Dobrescu:2021vak}, the interaction Lagrangian is  
\be
\label{eq:lagpre}
{\cal L}_{\rm int} = \frac{1}{2} V_\mu\left(  g_\chi  \bar \chi \gamma^\mu \chi + g_B  \sum_q   \frac{1}{3}  \bar q\gamma^\mu q\right),
\ee
where $g_B$ is the $U(1)_B$ gauge coupling, $q$ is a SM quark field,  and $g_\chi = g_B B_\chi$, where $B_\chi$ is the baryon number of $\chi$.  After spontaneous symmetry breaking, $V$ acquires a mass $m_V$. We assume that the scalar sector responsible for breaking $U(1)_B$ contains a scalar with baryon charge $-2B_\chi$, whose vacuum expectation value generates a small Majorana mass for $\chi$. After diagonalizing the full fermion mass matrix, the dark sector contains a pseudo-Dirac fermion with two mass eigenstates,
\be
\chi_1 = \frac{i(\xi - \eta)  }{\sqrt{2}}~~,~~
\chi_2 = \frac{\xi + \eta  }{\sqrt{2}},
\ee
with a small mass splitting $\delta \equiv m_2-m_1$.  In the mass basis, the interaction from Eq.~\eqref{eq:lagpre} becomes 
\be
{\cal L}_{\rm int}\supset \frac{i g_\chi}{2} V_\mu \bar \chi_1 \gamma^\mu \chi_2,
\ee
which realizes a version of inelastic dark matter \cite{Tucker-Smith:2001myb}.

Gauging $U(1)_B$ without adding new fermions leads to gauge anomalies, which we cancel by adding heavy fermions (``anomalons'') with suitable SM gauge quantum numbers and baryon charges. Following the prescription described in Ref. \cite{Dobrescu:2021vak}, we assume a set of anomalon fields such that the full theory satisfies the trace condition ${\rm Tr}(YB) = 0$, where $Y$ and $B$ are the hypercharge and baryon number matrices, and the trace runs over the SM quarks and anomalons. This ensures that the kinetic mixing between the hypercharge and $U(1)_B$ gauge bosons vanishes in the limit of equal fermion masses.

\begin{figure}[t!]
    \centering
    \includegraphics[width=\linewidth]{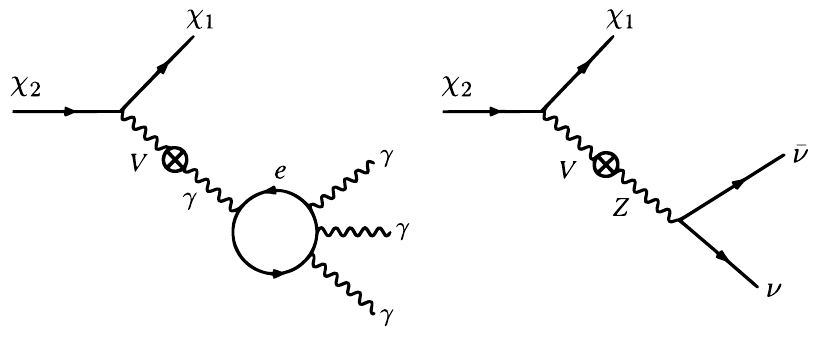}
    \caption{{\bf Left:} The Feynman diagram  for $\chi_2 \to \chi_1 + 3\gamma$ decays through $V\gamma$ kinetic mixing.  {\bf Right:} The Feynman diagram for $\chi_2 \to \chi_1 \bar\nu \nu$ decays through $VZ$ mass and kinetic mixing.  }
    \label{fig:decay}
\end{figure}

However, since quark and anomalon masses are not identical, $V\gamma$ and $VZ$ kinetic mixing will be induced, and additional $VZ$ mass mixing may also arise: 
\be
{\cal L}_{\rm mix} = \frac{1}{2}V_{\mu\nu}( \varepsilon_\gamma  F^{\mu\nu} + \varepsilon_Z Z^{\mu\nu}) + m_{V\! Z}^2  V_\mu Z^\mu,
\ee
where the kinetic mixing parameters are given in Ref.~\cite{Dobrescu:2021vak} and parametrically satisfy $\varepsilon_\gamma \sim \varepsilon_Z \sim e g_B /(16\pi^2)$. 
Within our regime of interest, if $\delta > 2m_e$, the process $\chi_2 \to \chi_1 e^+e^-$ is cosmologically prompt and unsuitable for maintaining a long-lived $\chi_2$ population. For $\delta < 2m_e$, however, the lifetime of $\chi_2$ can easily exceed the age of the universe. In this case, $V\gamma$ kinetic mixing  induces $\chi_2\to \chi_1 + 3\gamma$ decays through the electron loop shown in Fig.~\ref{fig:decay},  where for $\delta \ll m_\chi$~\cite{Pospelov:2008jk,Krnjaic:2025zjl},
\begin{align}
\label{eq:3gam}
\Gamma_{3\gamma}^{\rm KM} &= \frac{17 \varepsilon_\gamma^2 g_{\chi}^2  \alpha^4 }{
5472967500
\pi^5 } 
\frac{\delta^{13}}{m^8_e m^4_V} 
 \\
&\approx \frac{1}{
 10^{24} \, {\rm s} }  
\,\, \bigg(  \frac{ \varepsilon_\gamma g_\chi }{ 10^{-5}    } \bigg)^{ \! 2}  \! \bigg( \! \frac{\delta}{\rm  MeV} \! \bigg)^{\! \! 13}
\! \bigg( \frac{30 \, \rm GeV}{m_V} \bigg)^{\! 4} \! ,~
\end{align}
where $\alpha$ is the electromagnetic fine structure constant. In the same limit, $VZ$ kinetic mixing induces $\chi_2 \to \chi_1 \nu \bar \nu$ decays, where
\begin{align}
\Gamma^{\rm KM}_{ \nu\bar \nu}&= \frac{\varepsilon_Z^2 g^2 g^2_{\chi}}{5040 \pi^3 c^2_{W}} 
\frac{  \delta^9}{ m^4_Z m^4_V}
\\
&\approx \frac{1}{10^{32} {\rm s} }  \,
\bigg(  \frac{ \varepsilon_Z g_\chi }{ 10^{-5}    } \bigg)^{ \! 2}  
 \bigg( \! \frac{\delta}{ {\rm  MeV}} \! \bigg)^{\! \! 9}
\bigg( \frac{30 \, \rm GeV}{m_V} \bigg)^{\! 4}  ,~~~
\end{align}
where $g$ is the $SU(2)_L$ coupling and $c_W$ is the cosine of the weak mixing angle. These processes thus have a negligible impact on the $\chi_2$ population over the age of the universe.

The mass-mixing parameter, $m_{V\! Z}$, depends on the properties of the anomalons but can be negligible if these states acquire most of their mass from $U(1)_B$ breaking effects, and only small amounts from electroweak symmetry breaking~\cite{Dobrescu:2021vak}.

Neglecting kinetic mixing and taking the mass splitting to be much smaller than the DM mass, the mass-mixing contribution to the $\chi_2 \to \chi_1 \nu \bar \nu$ decay width is given by
\be
\Gamma^{\rm MM}_{\nu \bar \nu } \!
\!&=& \!\frac{g^2 g_{\chi}^2}{640\pi^3 c^2_{W}}
\frac{ m^4_{VZ}\delta^5}{m_Z^4 m_V^4} 
\\
&\approx& \!\! 
\frac{1}{
10^{18} \, {\rm s} }  \,  
\left(  \frac{g_\chi }{0.4} \right)^{ \! 2} \! \! \left( \! \frac{\delta}{ \rm MeV} \! \right)^{\! \! 5} 
\! \! \left( \frac{m_{V\! Z}}{10 \,  \rm MeV} \right)^{\! 4} 
\! \left( \frac{30 \, \rm GeV}{m_V} \right)^{\! \!4} \!\!
 ,~ ~~~~~~
\ee
 so for the remainder of this work, we will assume that $m_{VZ} \lesssim$ few MeV to ensure that  $\chi_2 \to \chi_1 \nu \bar \nu$ decays do not appreciably deplete the $\chi_2$ population over the age of the universe.

\medskip
\textbf{\textit{ Cosmology:}} In the early universe, the abundances of $\chi_1$ and $\chi_2$ evolve according to the following set of coupled Boltzmann equations:
\be
\frac{dn_{\chi_i}}{dt} 
+ 3H n_{\chi_i}
\! &=& \!
 - \langle \sigma v \rangle_{\chi_i \chi_i}  \big[ n_{\chi_i}^2 - (n_{\chi_i}^{\rm eq})^2 \big]  \nonumber \\  
&&\! -\langle \sigma v \rangle_{\chi_1 \chi_2}  \big[n_{\chi_1} n_{\chi_2}  - n_{\chi_1}^{\rm eq} n_{\chi_2}^{\rm eq} \big],~~ ~
\label{eq:boltzmann}
\ee
where $i=1,2$, $H$ is the Hubble rate, $n_{\chi_i}$ is the number density of species $\chi_i$, $n_{\chi_i}^{\rm eq}$ denotes that number density in equilibrium, and $ \langle \sigma v\rangle_{ij}$ is the thermally averaged annihilation/coannihilation cross section. Here, we are interested in the evolution of the total abundance, $n_{\chi} \equiv n_{\chi_1}+n_{\chi_2}$, allowing us to neglect conversion processes, $\chi_i f \leftrightarrow \chi_j f$. In the $\delta \ll T_f$ limit, where $T_f$ is the temperature at freeze-out, the evolution of $n_{\chi}$ can be written as
\begin{align}
\frac{d n_{\chi}}{dt}
+3 H n_{\chi}
=  -\langle \sigma_{\rm eff} v \rangle \big[n^2_{\chi}-(n^{\rm eq}_{\chi})^2 \big],
\end{align}
where $\langle \sigma_{\rm eff} v \rangle = (\langle \sigma_{\chi_1 \chi_1} v \rangle + \langle \sigma_{\chi_2 \chi_2} v \rangle +2 \langle \sigma_{\chi_1 \chi_2} v \rangle)/4$. Note that, throughout our analysis, we assume that the anomalons discussed earlier are decoupled during $\chi_i$ freeze-out and do not affect the DM relic abundance.

We focus here on the $m_\chi >m_V $ regime, in which annihilation into pairs of mediators is kinematically allowed. For the coupling hierarchy considered here, depletion of both DM species is dominated by $\chi_i\chi_i \to VV$, through the $t$-channel exchange of $\chi_j$, with $i\ne j$. The cross section for this process is given by~\cite{Pospelov_2009} 
\be
\langle \sigma v \rangle_{\chi_i \chi_i} \approx \frac{\pi \alpha_B^2 B_\chi^4}{16 m_\chi^2}
\frac{ (1 - m_V^2/m_\chi^2)^{3/2}}{  (1- m_V^2/2m_\chi^2)^2}~,
\ee
where $\alpha_B=g_B^2/4\pi$. This, in turn, corresponds to an effective cross section of $\langle \sigma_{\rm eff} v \rangle \approx \frac{1}{2}\langle \sigma v \rangle_{\chi_i \chi_i}$, where we have taken the non-relativistic limit. For $\langle \sigma_{\rm eff} v \rangle \approx 2 \times 10^{-26} \, {\rm cm}^3/{\rm s}$, the resulting thermal relic abundance is in good agreement with the total measured DM density, $\Omega_{\chi} h^2 \approx 0.12$.

\begin{figure}[t!]
    \centering
    \includegraphics[width=\linewidth]{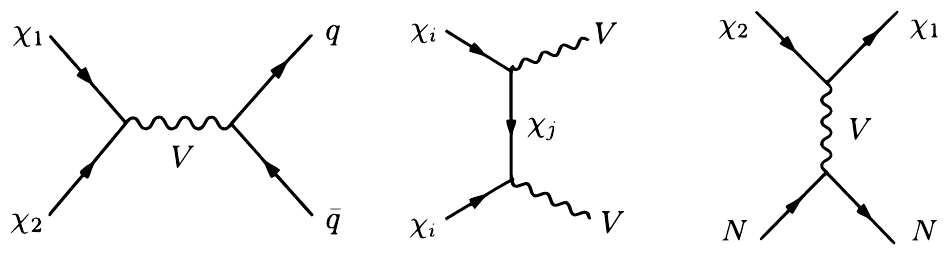}
    \caption{{\bf Left:} Feynman diagram for $\chi_1 \chi_2 \to q\bar q$ coannihilation, which is negligible in the parameter space under consideration here. {\bf Middle:} For $m_\chi > m_V$, $\chi_i \chi_i \to VV$ annihilation can proceed through $t$-channel $\chi_j$ exchange. {\bf Right:} Feynman diagram for exothermic downscattering to explain the LZ event, where $N$ is a xenon nucleus. Similar diagrams with $N$ replaced by $q$ govern $\chi_{1,2}$ kinetic decoupling from the SM in the early universe. }
    \label{fig:feyn}
\end{figure}

Even after $\chi_{1,2}$ chemically decouple from the SM at $T \sim m_\chi/20$, the large number density of relativistic SM particles can maintain efficient $\chi_i q \to  \chi_j q$ scattering through the process depicted in the rightmost diagram of Fig.~\ref{fig:feyn}. Below the QCD confinement scale, quarks are bound into hadrons, and scattering on baryons and antibaryons replaces scattering on free quarks. As the temperature falls below the baryon masses, the thermal baryon–antibaryon population becomes Boltzmann suppressed, eventually leaving a residual baryon density set by the cosmological baryon asymmetry. The resulting decrease in the scattering rate leads to the decoupling of these state-conversion processes. If decoupling occurs at a temperature much greater than $\delta$, the equilibrium suppression of the heavier state is negligible, and the two species have approximately equal number densities at that time. At $T\sim 100 \, {\rm MeV}$, 
\be
\frac{\Gamma_{\chi_i N \to \chi_j N}}{H} \sim 0.02
\, \brac{g_\chi g_B}{10^{-4}}^2   \brac{30 \, \rm GeV}{m_V}^4\!\! ,~~
\ee
where we have included thermally populated protons, neutrons, and their antiparticles. For the reference parameters, nucleon-mediated state conversions are therefore already inefficient at $T=100 \, {\rm MeV}$, well above the mass splitting considered here. There is thus no thermodynamic preference for 
$\chi_{1,2}$ states, and they are produced with nearly identical cosmological densities.

Similar conclusions apply to scattering on electrons, which is suppressed by kinetic mixing. Although  
$\chi_i e \to \chi_j e$ is not dramatically affected by hadronic confinement,  and $n_e \sim T^3$ down to  MeV-scale temperatures, 
the extra suppression from $\varepsilon_\gamma$ yields a kinetic-decoupling temperature of order 
\begin{align}
\label{eq:Tkd}
T_{\chi e}^{\rm kd} &\sim \bigg[\frac{m_{\chi} m_V^4}{(\varepsilon_{\gamma} e g_{\chi} )^2 M_{\rm Pl}}\bigg]^{1/4} \\
&\sim {\rm GeV} \,\,\bigg(\frac{10^{-5}}{g_{\chi} \varepsilon_{\gamma}}\bigg)^{1/2} \, \bigg(\frac{m_{\chi}}{35 \, {\rm GeV}}\bigg)^{1/4} \, \bigg(\frac{m_V}{30 \, {\rm GeV}}\bigg)~,\nonumber
\end{align}
so we can safely conclude that kinetic decoupling occurs well before temperatures of order $\delta$.

Although $\chi$-$e$ scattering becomes inefficient above the QCD confinement scale, scattering on quarks can continue to maintain kinetic equilibrium between DM and the SM plasma. Because quarks and leptons remain in thermal equilibrium with one another, efficient DM–quark scattering is sufficient to maintain a common temperature for the DM and SM populations. Provided that state-conversion processes maintain relative chemical equilibrium until a temperature $T_{\rm conv}\gg\delta$, the two states have approximately equal abundances when these processes decouple. The relative abundances of the two DM states are instead determined by the persistence of state-conversion processes. The population ratio at conversion freeze-out is
\be
\frac{n_2}{ n_1} = e^{-\delta/T_{\rm conv}} \approx 1~.
\ee
The two states therefore have approximately equal number densities at that time, and this ratio persists to the present day, provided that subsequent decays do not appreciably deplete the excited state.

\medskip
\textbf{\textit{  The Galactic Center Excess:}} It has long been appreciated that the spectrum and intensity of the GCE could be accommodated by models in which the DM annihilates to a pair of particles that then decay to SM quarks~\cite{Boehm:2014bia,Ko:2014gha,Abdullah:2014lla,Berlin:2014pya,Martin:2014sxa,Elor:2015tva,Escudero:2017yia,Hooper:2019xss,Hu:2025thq,Casas:2017rww,Hooper:2025fda}. Depending on the mass of the annihilation products and on their branching fractions to various final states, such scenarios typically favor DM masses in the range of $\sim 20$-$100 \, {\rm GeV}$ and annihilation cross sections near those required to obtain the observed thermal relic abundance. In the case of the baryon portal, as we are considering here, a gamma-ray spectrum that peaks near that of the GCE can be generated for $m_{\chi} \sim 20-100 \, {\rm GeV}$ and $m_V\sim 5-100 \, {\rm GeV}$ (see, for example, Fig.~7 of Ref.~\cite{Hooper:2019xss}).

\begin{figure}[t!]
    \centering
\includegraphics[width=0.98\linewidth]{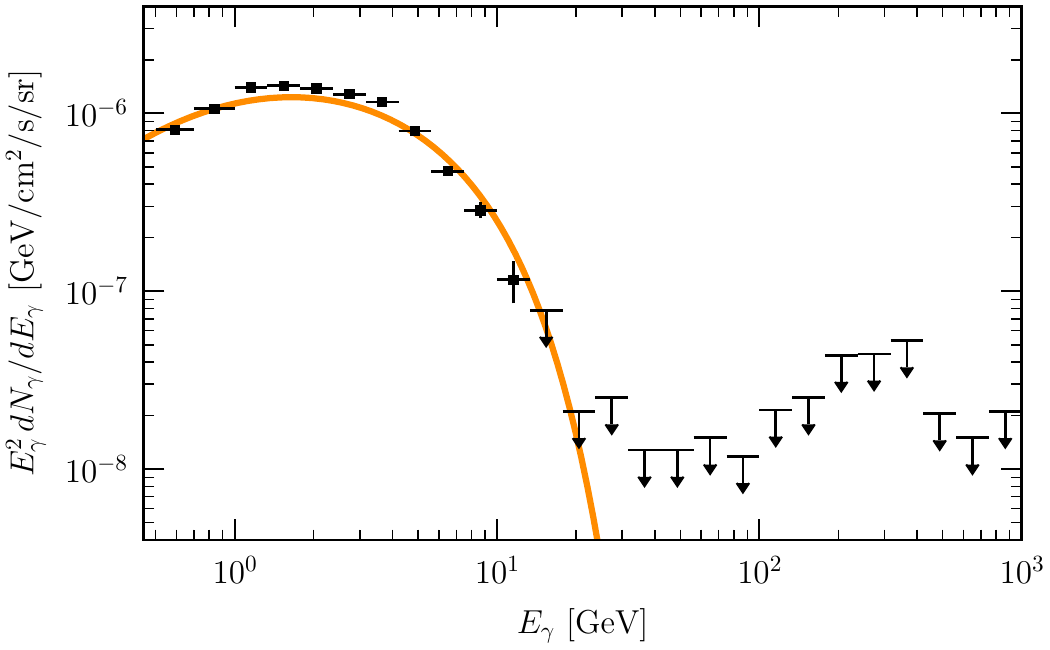}
    \caption{
    The spectrum of the gamma-ray emission generated from DM annihilations in our model, integrated within a $10^{\circ}$ radius around the Galactic Center. Here we have adopted $m_{\chi}=35 \, {\rm GeV}$, $m_V=30 \, {\rm GeV}$, $\sigma_{\chi_1 \chi_1} v= \sigma_{\chi_2 \chi_2} v= 4.4 \times 10^{-26} \, {\rm cm}^3/{\rm s}$ (which yields a thermal relic abundance equal to the measured DM density), and a generalized NFW DM halo profile with an inner slope of $\gamma=1.2$, a scale radius of 30 kpc, and normalized to a local density of 0.3 GeV/cm$^3$. This is compared to the measured spectrum of the Galactic Center Gamma-Ray Excess, as reported in Ref.~\cite{DiMauro:2026fnp}.}
    \label{fig:GCEspec}
\end{figure}

The DM annihilation rate per unit volume in this model can be written as
\begin{align}
\frac{d \Gamma_{\rm ann}}{dV} &= \frac{\langle \sigma_{\chi_1 \chi_1} v\rangle \, n^2_{\chi_1}}{2} + \frac{\langle \sigma_{\chi_2 \chi_2} v\rangle \, n^2_{\chi_2}}{2} \nonumber \\
& \approx  \frac{\pi \alpha_B^2 B_\chi^4 n^2_{\chi}}{64 m_\chi^2}\frac{ (1 - m_V^2/m_\chi^2)^{3/2}}{  (1- m_V^2/2m_\chi^2)^2},
\end{align}
where again $n_{\chi} \equiv n_{\chi_1} + n_{\chi_2}$. 

In Fig.~\ref{fig:GCEspec}, we show the spectrum of the gamma-ray emission generated from DM annihilations in our model, integrated within a $10^{\circ}$ radius around the Galactic Center. Here we have adopted $m_{\chi}=35 \, {\rm GeV}$, $m_V=30 \, {\rm GeV}$, and $\langle \sigma_{\chi_1 \chi_1} v \rangle= \langle \sigma_{\chi_2 \chi_2} v \rangle= 4.4 \times 10^{-26} \, {\rm cm}^3/{\rm s}$, which yields a thermal relic abundance in good agreement with the measured DM density. We will take this parameter point as a representative example of our model's ability to produce a GCE-like spectrum.

For the DM distribution, we have adopted a generalized Navarro-Frenk-White (NFW) profile with the following form:
\begin{align}
\rho_{\chi} \propto r^{-\gamma}  \, \big[1+(r/r_s)^2\big]^{\frac{\gamma-3}{2}},
\end{align}
where $r$ is the distance to the Galactic Center, $\gamma$ is the inner slope, and $r_s$ is the scale radius of the profile. The gamma-ray spectrum predicted in this model is in good agreement with the measured properties of the GCE. We note that there has been significant variance among past determinations of the GCE spectrum (see for example, Refs.~\cite{Daylan:2014rsa,Calore:2014xka,Fermi-LAT:2015sau,Cholis:2021rpp,DiMauro:2021raz}), which could potentially accommodate a relatively wide range of DM and mediator masses.

\medskip
\textbf{\textit{  Direct Detection and the LZ Event:}} The cross section for exothermic $\chi_2 \to \chi_1$ downscattering off a nuclear target can be written \cite{Lang:2010cd,deLima:2026shq}
\be
\frac{d\sigma}{dE_R} = 
\frac{\pi }{2v^2}
\frac{ (A B_\chi \alpha_B)^2  m_N  }{  (2m_N E_R + m_V^2 )^2 } F^2(q^2),
\ee
where $E_R$ is the nuclear recoil energy, $m_N$ and $A$ are the mass and mass number of the xenon isotope, $v$ is the relative velocity, $F(q^2)$ is the Helm form factor 
\cite{Helm:1956zz,LEWIN199687}, and $q = \sqrt{2m_N E_R}$ is the momentum transfer to the nucleus. Accounting for the velocity averaging, the event rate per unit detector mass is 
\be
\frac{dR}{dE_R} =  N_T \frac{\rho_{\chi _2}}{m_\chi}
\frac{ \pi (A B_\chi \alpha_B)^2  m_N  }{ 2(2m_N E_R + m_V^2 )^2 } F^2(q^2) \eta(v_{\rm min}),
\ee
where $N_T$ is the number of xenon nuclei per unit detector mass, $\rho_{\chi_2}$ is the local $\chi_2$ density, taken to be $\rho_{\chi_2} = 0.5\times 0.3$ GeV/cm$^3$, and $\eta$ is the mean inverse speed~\cite{McCabe:2010zh}
\be
\eta(v_{\rm min}) =  \int_{v_\mathrm{min}}^\infty \frac{d^3v}{v} f(v),
\ee
where $f(v)$ is the local DM velocity distribution in the standard halo model with $v_0=238$ km/s and $v_{\rm esc} = 544$ km/s \cite{Freese:2012xd}. The minimum speed for a given $E_R$ is 
\be
v_{\rm min}(E_R, \delta) = \frac{1}{\sqrt{2m_NE_R}}
\left| \,
\frac{m_N E_R}{\mu} - \delta
\,\right|~,
\ee
where $\mu$ is the $\chi$-$N$ reduced mass.

\begin{figure}[t!]
    \centering
\includegraphics[width=\linewidth]{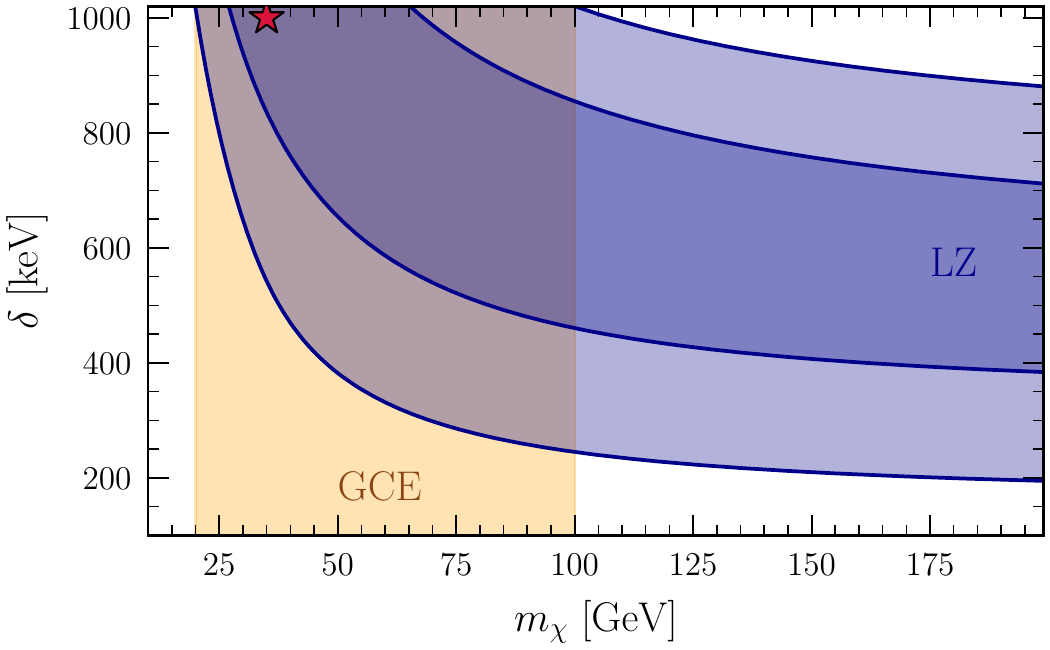}
    \caption{The blue contours enclose regions with $2\Delta(-\mathrm{ln}\mathcal{L}) <$ 2.30 and 6.18 in the fit to the single LZ event and the empty high-energy sideband. We profile over $m_V$ and $g_B$, with $g_\chi$ chosen to reproduce the observed relic abundance. The red star indicates our benchmark parameter point and the orange band indicates the DM mass range that could accommodate the GCE~\cite{Hooper:2019xss}.
    }
    \label{fig:LZcontours}
\end{figure}

To identify the parameter region favored by LZ, we construct an unbinned likelihood for a single event in the signal region, detected at $E_R=248$ keV, and no events in the high-energy sideband, which we take to be the 350–675 keV window: 
\be
-\mathrm{ln}\mathcal{L} = N_{\mathrm{ROI}} + N_{\mathrm{SB}} -\mathrm{ln} \:p(248 \mathrm{ keV}),
\ee
where $N_{\mathrm{ROI}}$ is the number of expected events in the signal region, $N_{\mathrm{SB}}$ is the number of expected events in the high-energy sideband, and $p(E_R)=dR/dE_R$. These values are calculated using the detector efficiency and resolution from Ref. \cite{LZ:2026axp}. We define our likelihood difference $2\Delta(-\mathrm{ln}\mathcal{L})$, with respect to the global fit minimum.

In Fig.~\ref{fig:LZcontours}, we show 2D contours in the ($m_\chi$, $\delta$) plane, corresponding to $2\Delta(-\mathrm{ln}\mathcal{L})=2.30,6.18$. At each ($m_\chi$, $\delta$) parameter point, the free parameters $(m_V,g_B)$ have been profiled over, while $g_{\chi}$ is fixed at the value such that the correct relic abundance is achieved. We show the contours in the ($m_\chi$, $\delta$) plane because these parameters control the minimum-speed threshold and therefore strongly influence the shape of the recoil spectrum. 

We find that a large range of $m_\chi$ values can accommodate the LZ event, although the global fit favors $m_\chi\sim50$ GeV and $\delta\sim1000$ keV. We restrict $\delta<2m_e$ so that $\chi_2$ is cosmologically stable. We also highlight the range of DM masses that could accommodate the GCE, as presented in Ref.~\cite{Hooper:2019xss}, finding a large mass window in which the regions overlap.

\begin{figure}[t!]
    \centering
\includegraphics[width=\linewidth]{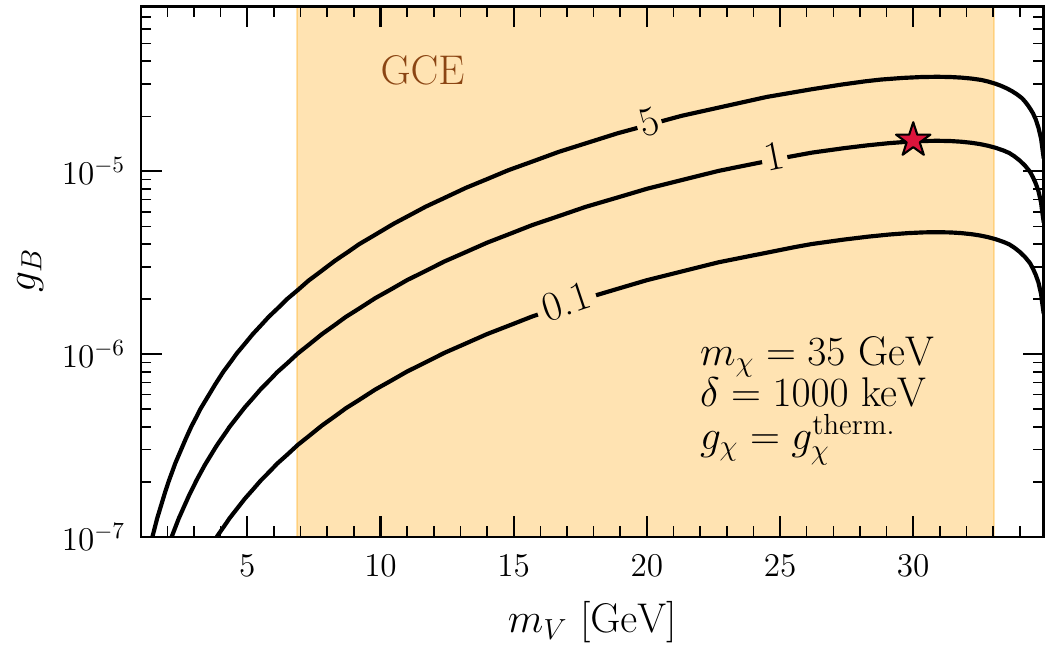}
    \caption{
    Contours of the expected number of events at LZ in the signal region. The red star indicates our benchmark parameter point and the orange band indicates the $m_V$ range which could accommodate the GCE for $m_\chi=35$ GeV~\cite{Hooper:2019xss}. For each point in this plane we choose $g_\chi$ to obtain the observed relic density. }
    \label{fig:NEvents}
\end{figure}

To explore the remaining parameter dependence, we fix $m_\chi=35 \,{\rm GeV}$ and $\delta=1000 \, {\rm keV}$, corresponding to the red star in Fig.~\ref{fig:LZcontours}. This choice lies within the region favored by the LZ fit and permits mediator masses that can accommodate the GCE. Fig.~\ref{fig:NEvents} shows the expected LZ signal count as a function of $m_V$ and $g_B$, with $g_\chi$ chosen at each point to reproduce the observed relic abundance. The orange band indicates the mediator-mass range compatible with the GCE for this DM mass, as presented in Ref.~\cite{Hooper:2019xss}. Within this band, $g_B \sim 10^{-6}–10^{-5}$ yields approximately one expected LZ event, depending on the mediator mass.

\begin{figure}[t!]
    \centering
    \includegraphics[width=\linewidth]{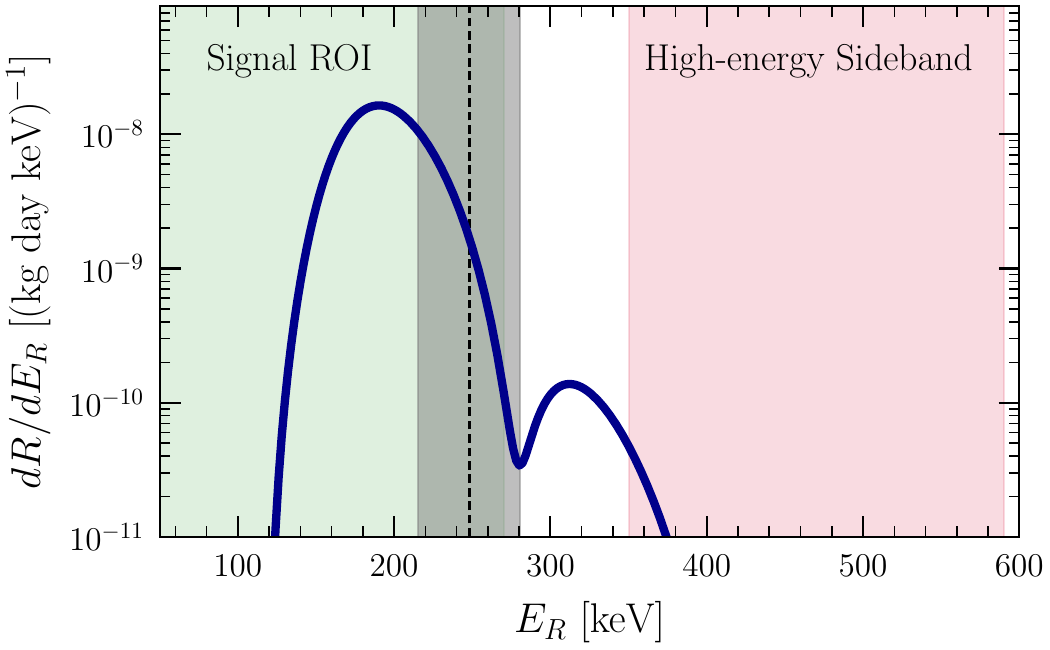}
    \caption{Recoil spectra for our benchmark point ($m_\chi=35 \,{\rm GeV}$, $m_V=30 \, {\rm GeV}$, $\delta=1000 \, {\rm keV}$, $g_B=1.46\times10^{-5}$ and $g_{\chi}=0.32$). The green band indicates the signal recoil energy window, while the red band shows the range of the high-energy sideband. The LZ event recoil energy is shown in black, with a shaded region indicating the uncertainty, where the statistical and systematic errors have been added in quadrature.
    }
    \label{fig:LZspec}
\end{figure}

In Fig.~\ref{fig:LZspec} we show the predicted recoil spectra of our benchmark parameter point. The green and red bands indicate the signal region and high-energy sideband, respectively. The reconstructed energy of the LZ event is marked in black, and the shaded region shows its uncertainty, obtained by adding the statistical and systematic uncertainties in quadrature. This benchmark yields an expected number of events of $N_{\mathrm{ROI}}=1$ and $N_{\mathrm{SB}}=8\times10^{-4}$, and thus illustrates this model's ability to provide a good fit to the single LZ event and the GCE (as shown in Fig.~\ref{fig:GCEspec}), while yielding the observed DM relic abundance.

\medskip
\textbf{\textit{Collider Bounds: }} Our scenario contains both a quark-coupled gauge boson, $V$, and SM-charged anomalons, whose properties can be constrained by collider searches. The hadrophilic vector decays via $V \to \bar q q$, yielding dijet resonance events at high-energy collider experiments. Low-mass dijet searches by ATLAS~\cite{ATLAS:2024bms} and CMS~\cite{CMS:2019xai,CMS:2026yvw} place upper limits of $g_B \lesssim 0.1-0.3$ over the $m_V \sim 10 -200 \,{\rm GeV}$ mass range. For $m_V \lesssim m_Z$, earlier bounds from UA1, UA2, LEP, and Tevatron searches for dijet resonances constrain $g_B \lesssim {\cal O}(1)$ \cite{Krnjaic:2011ub,Dobrescu:2013cmh}. 
Since we only require $g_B \sim 10^{-5}$ to explain the LZ excess, these constraints can easily be accommodated within the favored parameter space of our model. 

The anomalons required for anomaly cancellation are charged under both $U(1)_B$ and SM hypercharge, and are thus subject to collider constraints. Since there is no unique set of anomalons, these limits will vary across realizations and a comprehensive study of these possibilities is beyond the scope of this work. 
However, since the anomalons $\{\psi\}$ are chiral under $U(1)_B$, their masses arise from its spontaneous symmetry breaking. In a simple realization involving a single Higgs-like field $\phi$, the anomalon masses satisfy $m_\psi \sim y_\psi v_\phi \sim y_\psi m_V/g_B$, so we 
obtain~\cite{Kahn:2016vjr}
\be
m_\psi  
\sim 3 \, {\rm TeV} \,
\bigg(
\frac{y_\psi}{10^{-3}} \bigg) 
\bigg(\frac{m_V}{30 \, \rm GeV} \bigg)
\bigg(\frac{10^{-5}}{g_B} \bigg)
,
\ee
where $y_\psi$ is the anomalon Yukawa coupling to $\phi$, and $v_\phi$ is its expectation value.
Thus, depending on the value of $y_\psi$, the anomalons could be quite heavy, although scenarios in which the anomalons are accessible at colliders are also possible.

\medskip
\textbf{\textit{ Discussion:}} In summary, we have presented in this \textit{Letter} a simple model of inelastic DM coupled to a new $U(1)_B$ gauge boson that can simultaneously explain the LZ event and the Galactic Center Gamma-Ray Excess. The direct detection signal arises from exothermic $\chi_2 \to \chi_1$ downscattering off nuclei. Unlike models in which this process is mediated by a coupling to electric charge~\cite{deLima:2026shq}, our model predicts a coherent enhancement proportional to $A^2$ rather than $Z^2$. 
Furthermore, for a downscattering signal, the peak of the recoil distribution depends on the mass of the target nucleus and on the form factor for a given target. Thus, with sufficient data from experiments using different target materials, the peak shift and $A^2$ signal dependence could be used to distinguish this model from alternative scenarios.

\medskip
\textbf{\textit{Acknowledgments:}} CG and DH are supported by the Office of
the Vice Chancellor for Research at the University of
Wisconsin–Madison with funding from the Wisconsin
Alumni Research Foundation. This manuscript has been authored in part by Fermi Forward Discovery Group, LLC under Contract No.~89243024CSC000002 with the U.S. Department of Energy, Office of Science, Office of High Energy Physics. This work
was supported in part by the Kavli Institute for Cosmological Physics at the University of Chicago through an
endowment from the Kavli Foundation and its founder
Fred Kavli.

\bibliographystyle{apsrev4-1}
\bibliography{references}

\renewcommand{\theequation}{A-\arabic{equation}}
\setcounter{equation}{0}
\renewcommand{\thefigure}{A-\arabic{figure}}
\setcounter{figure}{0}

\end{document}